\documentclass[journal,10pt,onecolumn,draftclsnofoot]{IEEEtran}
\usepackage{url}
\usepackage{multirow}
\usepackage{longtable}
\usepackage{amssymb}
\usepackage{mathrsfs}
\usepackage{graphicx}
\usepackage{amsfonts}
\usepackage{amsmath}
\usepackage{hyperref}
\usepackage{algorithm}
\usepackage{algorithmic}
\usepackage{diagbox}
\usepackage{subfig}
\usepackage{threeparttable}
\usepackage{subfig}
\usepackage{framed}
\usepackage{color}

\ifCLASSINFOpdf
\else
\fi
\begin{document}
%
\title{A Comparison of Word Embeddings for the Biomedical Natural Language Processing}


\author{\IEEEauthorblockN{Yanshan Wang*, Sijia Liu, Naveed Afzal, Majid Rastegar-Mojarad, Liwei Wang, Feichen Shen, Paul Kingsbury, Hongfang Liu*\footnote{*Corresponding authors.}} \\
\IEEEauthorblockA{Department of Health Sciences Research\\
Mayo Clinic\\
Rochester, USA\\
Email: \{Wang.Yanshan, Liu.Sijia, Afzal.Naveed, Mojarad.Majid, Wang.Liwei, Shen.Feichen, Kingsbury.Paul1, Liu.Hongfang\}@mayo.edu} 
}


%


\maketitle

\begin{abstract}
\textbf{Background}
Word embeddings have been prevalently used in biomedical Natural Language Processing (NLP) applications due to the vector representations of words capturing useful semantic properties and linguistic relationships between words. Different textual resources (e.g., Wikipedia and biomedical literature corpus) have been utilized in biomedical NLP to train word embeddings and these word embeddings have been commonly leveraged as feature input to downstream machine learning models. However, there has been little work on evaluating the word embeddings trained from different textual resources. 

\textbf{Methods}
In this study, we empirically evaluated word embeddings trained from four different corpora, namely clinical notes, biomedical publications, Wikipedia, and news. For the former two resources, we trained word embeddings using unstructured electronic health record (EHR) data available at Mayo Clinic and articles (MedLit) from PubMed Central, respectively. For the latter two resources, we used publicly available pre-trained word embeddings, GloVe and Google News. The evaluation was done qualitatively and quantitatively. For the qualitative evaluation, we arbitrarily selected medical terms from three medical categories (i.e., disorder, symptom, and drug), and manually inspected the five most similar words computed by word embeddings for each of them. We also analyzed the word embeddings through a 2-dimensional visualization plot of 377 medical terms. For the quantitative evaluation, we conducted both intrinsic and extrinsic evaluation. For the intrinsic evaluation, we evaluated the medical semantics of word embeddings using four published datasets for measuring semantic similarity between medical terms, i.e., Pedersen's dataset, Hliaoutakis's dataset, MayoSRS, and UMNSRS. For the extrinsic evaluation, we applied word embeddings to multiple downstream biomedical NLP applications, including clinical information extraction (IE), biomedical information retrieval (IR), and relation extraction (RE), with data from shared tasks.

\textbf{Results}
The qualitative evaluation shows that the word embeddings trained from EHR and MedLit can find more relevant similar medical terms than those from GloVe and Google News. The intrinsic quantitative evaluation verifies that the semantic similarity captured by the word embeddings trained from EHR is closer to human experts' judgments on all four tested datasets. The extrinsic quantitative evaluation shows that the word embeddings trained on EHR achieved the best F1 score of 0.900 for the clinical IE task; no word embeddings improved the performance for the biomedical IR task; and the word embeddings trained on Google News had the best overall F1 score of 0.790 for the RE task. 

\textbf{Conclusion}
Based on the evaluation results, we can draw the following conclusions. First, the word embeddings trained on EHR and MedLit can capture the semantics of medical terms better and find semantically relevant medical terms closer to human experts' judgments than those trained on GloVe and Google News. Second, there does not exist a consistent global ranking of word embeddings for all downstream biomedical NLP applications. However, adding word embeddings as extra features will improve results on most downstream tasks. Finally, the word embeddings trained on biomedical domain corpora do not necessarily have better performance than those trained on general domain corpora for any downstream biomedical NLP task. 

\end{abstract}


%
\IEEEpeerreviewmaketitle

\section{Introduction}

Word embeddings have been prevalently used in Natural Language Processing (NLP) applications due to the vector representations of words capturing useful semantic properties and linguistic relationships between words using deep neural networks \cite{mikolov2013linguistic, liu2016learning, levy2014dependency}. Word embeddings are commonly utilized as feature input to machine learning models, which enables machine learning techniques to process rew text data. There has been an increasing number of studies applying word embeddings in common NLP tasks, such as information extraction (IE) \cite{wang2017clinical, zeng2014relation, nguyen2014employing}, information retrieval (IR) \cite{ganguly2015word}, sentiment analysis \cite{tang2014learning,maas2011learning}, question answering \cite{ren2015exploring,dong2015question}, and text summarization \cite{yogatama2015extractive, rush2015neural}. Recently, in the biomedical domain, word embeddings have been remarkably utilized in applications such as biomedical named entity recognition (NER) \cite{tang2014evaluating, liu2015effects}, medical synonym extraction \cite{jagannatha2015mining}, relation extraction (RE) (e.g., chemical-disease relation \cite{jiang2015crd}, drug-drug interaction \cite{liu2016drug,wang2017dependency} and protein-protein interaction \cite{jiang2016general}), biomedical IR \cite{jo2016cbnu, wang2016ensemble} and medical abbreviation disambiguation \cite{wu2015clinical}. 

There are two main text resources utilized to train word embeddings for the biomedical NLP applications: internal task corpora (e.g., training data) \cite{jo2016cbnu} and external data resources (e.g., Wikipedia) \cite{gurulingappa2016semi}. The use of the former resource is straightforward as the internal corpora capture the nuances of language topic specific to the task \cite{diaz2016query}. Exploiting external data resources is based on an implicit assumption that the external resources contain knowledge that could be used to enhance domain tasks \cite{shen2016knowledge,shen2016predicate,shen2015bmqgen}.  In addition, a number of pre-trained word embeddings are publicly available, such as the embeddings of Google News\footnote{https://drive.google.com/file/d/0B7XkCwpI5KDYNlNUTTlSS21pQmM/edit} and GloVe\footnote{https://nlp.stanford.edu/projects/glove/}. These embeddings could capture the semantics of general English words from a large corpus. However, one question remains unanswered: do we need to train word embeddings for a specific NLP task given a number of public pre-trained word embeddings? This question becomes more significant for domain areas, and particularly more important for the clinical domain. The reason is that little electrical health record (EHR) data is publicly available due to the Health Insurance Portability and Accountability Act (HIPAA) requirements, while biomedical literature is more widely available through resources such as PubMed\footnote{https://www.ncbi.nlm.nih.gov/pubmed/}. However, to the best of our knowledge, there has been little work done to evaluate word embeddings trained from these textual resources for biomedical NLP applications.

In this study, we empirically evaluated word embeddings trained from four different corpora, namely clinical notes, biomedical publications, Wikipedia, and news. For the former two resources, we utilized clinical notes from the EHR system at Mayo Clinic and articles from PubMed Central (PMC)\footnote{http://www.ncbi.nlm.nih.gov/pmc/tools/openftlist/} to train word embeddings. For the latter two resources, we used publicly available pre-trained word embeddings, GloVe and Google News. We performed the evaluation qualitatively and quantitatively. For the qualitative evaluation, we adopted the method used in Levy and Goldberg's study \cite{levy2014dependency}, and manually inspected five of the most similar medical words to each of arbitrarily selected medical words from three medical categories (disorder, symptom, and drug). In addition, we analyzed word embeddings through a 2-dimensional visualization plot of 377 medical words. For the quantitative evaluation, we conducted both intrinsic and extrinsic evaluation. The intrinsic evaluation directly tested semantic relationships between medical words using four published datasets for measuring semantic similarity between medical terms, i.e., Pedersen \cite{pedersen2007measures}, Hliaoutakis \cite{hliaoutakis2005semantic}, MayoSRS \cite{pakhomov2011towards}, and UMNSRS \cite{pakhomov2010semantic,pakhomov2016corpus}. For the extrinsic evaluation, we applied word embeddings to downstream NLP applications in the biomedical domain including clinical IE, biomedical IR, and RE, and measured the performance of word embeddings. 

\section{Related Work}

Due to the success of word embeddings in a variety of NLP applications, some existing studies evaluate word embeddings in representing word semantics quantitatively. Most of them  focus on evaluating the word embeddings generated by different approaches. Baroni et al. \cite{baroni2014don} presented the first systematic evaluation of word embeddings generated by four models, i.e., DISSECT\footnote{http://clic.cimec.unitn.it/composes/}, CBOW \cite{mikolov2013linguistic} using word2vec\footnote{https://code.google.com/p/word2vec/}, Distributional Memory model\footnote{http://clic.cimec.unitn.it/dm/}, and 4) Collobert and Weston model\footnote{http://ronan.collobert.com/senna/} using a corpus of 2.8 billion tokens in the general English domain. They tested these models on fourteen benchmark datasets in five categories, including semantic relatedness, synonym detection, concept categorization, selectional preferences, and analogy. They found that the word2vec model, CBOW, performed the best for almost all the tasks. 

Schnabel et al. \cite{schnabel2015evaluation} trained the CBOW model of word2vec \cite{mikolov2013linguistic}, C\&W embeddings \cite{collobert2011natural}, Hellinger PCA \cite{lebret2013word}, GloVe \cite{pennington2014glove}, TSCCA \cite{dhillon2012two}, and Sparse Random Projections \cite{li2006very} on a 2008 GloVe dump, and tested on the same fourteen datasets. They found that the CBOW outperformed other embeddings on 10 datasets. They also conducted an extrinsic evaluation by using the embeddings as input features to two downstream tasks, namely noun phrase chunking and sentiment classification. They found the results of CBOW were also among the best. 

Ghannay et al. \cite{ghannay2016word} conducted a similar intrinsic evaluation, they additionally evaluated the skip-gram models of word2vec \cite{mikolov2013linguistic}, CSLM word embeddings \cite{schwenk2013cslm}, dependency-based word embeddings \cite{levy2014dependency}, and combined word embeddings on four NLP tasks, including Part-Of-Speech tagging, chunking, named entity recognition, mention detection, and two linguistic tasks. They trained these word embeddings on the Gigaword corpus composed of 4 billion words and found that the dependency-based word embeddings gave the best performance on the NLP tasks and that the combination of embeddings yielded significant improvement. Nayak et al's study \cite{nayak2016evaluating} recommended that the evaluation of word embeddings should test both syntactic and semantic properties, and that the evaluation tasks should be closer to real-word applications. However, few of these studies evaluated word embeddings for tasks in the biomedical domain.

As most of the aforementioned studies evaluate word embeddings in the general (i.e., non-biomedical) NLP domain, only one recent study by Pakhomov et al. \cite{pakhomov2016corpus} evaluates word embeddings in the biomedical domain, to the best of our knowledge. They trained the CBOW model on two biomedical corpora, namely clinical notes and biomedical publications, and one general English corpora, namely GloVe. The word embeddings were evaluated on subsets of UMNSRS dataset, which consisted of pairs of medical terms with the similarity of each pair assessed by medical experts, and on a document retrieval task and a word sense disambiguation task. They found that the semantics captured by the embeddings computed from biomedical publications were on par with that from clinical notes. We extended their evaluation of word embeddings by: 1) utilizing four datasets to evaluate word embeddings on capturing medical term semantics; 2) conducting a qualitative evaluation; and 3) examining word embeddings on more downstream applications with data provided by shared biomedical NLP tasks.

\section{Word Embeddings and Parameter Settings}
We utilized word2vec in this study as it has been shown that word2vec generates better word embeddings for most general NLP tasks than other approaches \cite{baroni2014don,schnabel2015evaluation}. Since no evidence shows that the CBOW architecture outperforms the skip-gram architecture or vice versa, we arbitrarily chose the skip-gram architecture for word2vec.


Word embeddings can be represented as a mapping $V\rightarrow \mathbb{R}^D: w\mapsto \theta$, which maps a word $w$ from a vocabulary $V$ to a real-valued vector $\theta$ in an embedding space with the dimension of $D$. The skip-gram architecture, proposed by Mikolov et al. \cite{mikolov2013linguistic}, uses the focus word as the single input layer, and the target contextual words as the output prediction layer. To avoid expensive computation over every word in $V$, Mikolov et al. \cite{mikolov2013linguistic} proposed a technique called ``negative-sampling'' that samples a few output words and updates embeddings for this small sample in each iteration. We formulate the model mathematically in the following. Given a sequence of target word $w_1, w_2, ..., w_T$ and its contextual word $h_1, h_2, ..., h_T$, the training objective is to maximize the conditional log probability of observing the actual output contextual word given the input target word, i.e.,
\begin{equation}\label{equ.obj}
\max J = \max \frac{1}{T}\sum_{t=1}^T \log P(h_t|w_t).
\end{equation}
where $J$ is the objective function, and $P(h|w)$ is the conditional probability in the neural probabilistic language model. $P(h|w)$ is usually defined by
\begin{equation}
P(h|w)=\frac{e^{\theta_h^\mathsf{T}\theta'_w}}{\sum_{h\in V} e^{\theta_h^\mathsf{T}\theta'_w}},
\end{equation}
where $\theta'$ and $\theta$ are the input and output word embeddings, respectively. Accordingly, the log probability can be written as:
\begin{equation}\label{equ.log}
\log P(h|w) = \theta_h^\mathsf{T}\theta'_w-\log(\sum_{h\in V}e^{\theta_h^\mathsf{T}\theta'_w}).
\end{equation}
We can take the derivative of $J$ to obtain the embeddings, updating the equation iteratively. However, the computation is extremely expensive as in each iteration, the algorithm needs to go through the vocabulary $V$. By using negative-sampling, Mikolov et al. \cite{mikolov2013linguistic} defined an empirical log probability $P'(h|w)$ to approximate $P(h|w)$:
\begin{equation}\label{equ.log.mod}
P'(h|w) =\log \sigma(\theta_h^\mathsf{T}\theta'_w)+\sum_i^k \mathbb{E}_{h\sim P_n(h)} [ \log\sigma(-\theta_h^\mathsf{T}\theta'_w)],
\end{equation}
where $\sigma(x)=1/(1+\exp(-x))$ is a softmax function that normalizes a real vector into a probability vector, $P_n(h_i)=\frac{f(h_i)^{3/4}}{\sum_i^{|V|} f(h_i)^{3/4}}$ is an empirical distribution that generates $k$ negative samples with $f(h_i)$ being the term frequency for term $h_i$. The word embeddings $\theta$ can be computed by maximizing the objective function in Equation (\ref{equ.obj}) by replacing $P(h|w)$ with $P'(h|w)$. 


We tested different vector dimensions of $D$ (i.e., 20, 60, 100) for the vector representation trained on EHR and MedLit and chose 100 for EHR and 60 for MedLit according to the performance in our intrinsic evaluation. Similarly, we chose the dimension of 100 for GloVe, and 300 for Google News since only 300 was publicly available for Google News. The experimental results of using different vector dimensions for the word embeddings are provided in Appendix \ref{ap.1}. For training word embeddings on the EHR and MedLit, we set the window size to 5, the minimum word frequency to 7 (i.e., the words that occurred less than 7 times in the corpus were ignored), and the negative sampling parameter to 5. These parameters were selected based on previous studies \cite{mikolov2013linguistic,levy2014dependency,wang2017dependency}.

\section{Data and Text Pre-prosessing}


The first corpus, denoted as EHR, contains textual clinical notes for a cohort of 113k patients receiving their primary care at Mayo Clinic, spanning a period of 15 years from 1998 to 2013. The vocabulary size of this corpus is 103k. The second corpus, denoted as MedLit, is obtained from a snapshot of the Open Access Subset\footnote{http://www.ncbi.nlm.nih.gov/pmc/tools/openftlist/} of PubMed Central (PMC)\footnote{http://www.ncbi.nlm.nih.gov/pmc/} in March 2016, which is an online digital database of freely available full-text biomedical literature. It contains 1.25 million biomedical articles, and 2 million distinct words in the vocabulary. As a comparison, additional public pre-trained word embeddings from two general English resources, i.e., Google News  \footnote{https://drive.google.com/file/d/0B7XkCwpI5KDYNlNUTTlSS21pQmM/edit?usp=sharing} and GloVe \footnote{http://nlp.stanford.edu/data/glove.6B.zip}, were utilized in the evaluation. The Google News embeddings have vector representations for 3 million words from Google News, trained by the word2vec \cite{mikolov2013linguistic}. The GloVe embeddings were trained by the GloVe model \cite{pennington2014glove}, and have 400k unique words in the vocabulary from a snapshot of Wikipedia in 2014 and Gigaword Fifth Edition\footnote{https://catalog.ldc.upenn.edu/LDC2011T07}. 


The MedLit and EHR corpora were pre-processed minimally by removing punctuation, lowercasing, and replacing all digits with "7". One exception is that we replaced `-' with `\_' if two or more words were connected by `-' and treated these words as one. For the MedLit corpus, we additionally removed website urls, email addresses, and twitter handles. For the EHR corpus, the clinical narratives are written by medical practitioners, and thus contain more incomplete sentences than research articles. Therefore, we conducted additional pre-processing on the EHR corpus specific for the clinical notes from Mayo Clinic. Specifically, the section of ``Family history" in the corpus was removed if it was semi-structured \cite{wang2017systematic}. As shown by an example in Table \ref{tab.family}, the semi-structured ``Family history'' section does not provide much valuable semantic information. The section of ``Vital Signs'' was totally removed since it did not contain contextual information for training word embeddings. Table \ref{tab.vital} shows an example of the ``Vital Signs'' section in the EHR corpus. Moreover, we replaced all text contractions with their respective complete text (e.g., ``can't'' is replaced with ``can not''), and removed all the clinical notes metadata and note section headers, dates, phone numbers, weight and height information, and punctuation.

\begin{table}[!h]
\centering
\footnotesize
\caption{An example of the semi-structured ``Family history'' section from the EHR corpus.}
\label{tab.family}
\begin{tabular}{l}
\hline
\textit{MOTHER} \\
  \textit{Stroke/TIA} \\
  \textit{BROTHERS} \\
  \textit{4 brothers alive  1 brother deceased} \\
  \textit{SISTERS} \\
  \textit{2 sisters alive} \\
  \textit{DAUGHTERS} \\
  \textit{1 daughter alive} \\
  \textit{Heart disease } \\
\hline
\end{tabular}
\end{table}

\begin{table}[!h]
\centering
\footnotesize
\caption{An example of the ``Vital Signs'' section from the EHR corpus.}
\label{tab.vital}
\begin{tabular}{c}
\hline
\textit{Height:  149.1 cm.  Weight:  44.5 kg.  BSA(G):  1.3573 M2.  BMI:  20.02 KG/M2.} \\
\hline
\end{tabular}
\end{table}



\section{Qualitative Evaluation}
We arbitrarily selected medical words from three medical semantic categories, namely disorder, symptom, and drug. Word embeddings trained from four different corpora were utilized to compute the five most similar words to each selected medical word according to the cosine similarity. Then we adopted the method used in Levy and Goldberg's study \cite{levy2014dependency} and manually inspected the conceptual similarity between the target word and the most similar words. Suppose $w_1$ and $w_2$ are two words, the similarity between $w_1$ and $w_2$ is defined as 
\begin{equation}\label{equ.simi1}
\text{similarity}(w_1, w_2)=\frac{\theta_1\cdot \theta_2}{\|\theta_1\|\|\theta_2\|},
\end{equation}
where $\theta_1$ and $\theta_2$ are vector representations for $w_1$ and $w_2$ in the embedding space, respectively. If the target word is a medical phrase $s_1$ consisting of multiple words, i.e., $s_1={w_1, w_2, ..., w_n}$, the similarity function becomes
\begin{equation}\label{equ.simi2}
\text{similarity}(s_1, w_2)=\frac{\Theta_1\cdot \theta_2}{\|\Theta_1\|\|\theta_2\|}
\end{equation}
where $\Theta_1=\frac{1}{n}\sum_i^n \theta_i$ is the representation for $s_1$ in the embedding space. This is different from Pakhomov et al's study \cite{pakhomov2016corpus} where only single word terms were considered. We ranked the words in the vocabulary based on the similarity to the target word and chose the five top ranked words.

Table \ref{tab.words} lists eight target words from the three medical categories, and the corresponding five most similar words computed by using the word embeddings trained from different resources. 

For the first target word describing a disorder, \textit{diabetes}, EHR and MedLit find its synonym, \textit{mellitus}, in the most similar words while GloVe and Google News fail to find it. EHR finds two terms related to co-morbidities of \textit{diabetes}, which are \textit{cholesterolemia} and \textit{dyslipidemia}, and a common adjective modifier term, \textit{uncontrolled}. MedLit finds terms relevant to co-existing conditions for \textit{diabetes}, such as \textit{cardiovascular} (possibly from \textit{cardiovascular disease}), \textit{nonalcoholic} (possibly from \textit{nonalcoholic fatty liver disease}), \textit{obesity}, and \textit{polycystic} (possibly from \textit{polycystic ovary syndrome} which is a hyperandrogenic disorder that is associated with a high-risk of development of Type 2 diabetes). Most of these terms are related with medical research topics and occur frequently in the biomedical research articles. GloVe finds two related terms, \textit{hypertension} and \textit{obesity}, while three other terms, i.e., \textit{arthritis}, \textit{cancer} and \textit{alzheimer}, are less relevant disease names. Google News finds two morphological terms, \textit{diabetics} and \textit{diabetic}, relevant to the target words, one synonym, \textit{diabetes\_mellitus}, and one related disease name, \textit{heart disease}. We can draw similar conclusions for the second and third disorder words.

The \textit{dyspnea} example in the symptom category demonstrates the advantage of EHR and MedLit. EHR finds \textit{palpitations}, a common cause of \textit{dyspnea}, and \textit{orthopnea}, \textit{exertional}, and \textit{doe} (dyspnea on exertion) are synonyms or specific conditions for \textit{dyspnea}. MedLit finds related symptoms, \textit{sweats} and \textit{orthopnea}, a synonym \textit{breathlessness}, a relevant disorder \textit{hypotension}, and a term relevant to the symptom \textit{rhonchi}. GloVe finds synonyms \textit{shortness} and \textit{breathlessness},  and less relevant symptoms \textit{cyanosis} and \textit{photophobia}. Google News finds less relevant symptoms \textit{pruritus} and \textit{rhinorrhea} and less relevant disease \textit{nasopharyngitis}. Similar observations can be found for \textit{sore throat} and \textit{low blood pressure} as well.

We can further observe that the semantics captured by the word embeddings trained from different corpora is disparate for the medical terms in the drug category. For \textit{opioid}, EHR finds \textit{opiate}, \textit{benzodiazepine}, \textit{sedative}, \textit{polypharmacy}, which are very relevant medications. MedLit finds \textit{nmda\_receptor}, \textit{affective\_motivational}, \textit{naloxone\_precipitated}, \textit{hyperlocomotion}, which are related to the mechanism of  action of \textit{opioid}. GloVe finds \textit{analgesic} and less relevant \textit{anti-inflammatory}, and Google News finds \textit{opioid}-related phrases and relevant term \textit{antipsychotics}. For the target term \textit{aspirin}, EHR also finds very clinically relevant used terms and MedLit finds relevant terms in research articles while GloVe and Google News only find medication names.

It is obviously shown from these target words and the corresponding similar words that EHR and MedLit can capture the semantics of medical terms better than GloVe and Google News and find more relevant similar medical terms. However, EHR and MedLit find similar medical terms from different perspectives due to their focus difference. EHR contains clinical narratives and thus it is closer to clinical language. It contains terms with different morphologies and even typos, such as \textit{melitis}, \textit{caner} and \textit{thraot} as listed in Table \ref{tab.words}. Differently, MedLit contains more medical terms used in research articles, and finds similar words mostly from a biomedical research perspective. 


\begin{table}
\centering
\footnotesize
\caption{Selected medical words from three medical semantic categories (i.e., disorder, symptom, and drug) and the corresponding five most similar words induced by the word embeddings trained from different resources.}
\label{tab.words}
\begin{tabular}{p{1.5cm}|p{2.3cm}p{2.2cm}p{2.2cm}p{2.2cm}p{2.2cm}}
\hline
Semantic Category & Target Word & EHR & MedLit & GloVe & Google News \\
\hline
\multirow{3}{*}{Disorder}  & diabetes & mellitus, \newline uncontrolled, \newline cholesterolemia, \newline dyslipidemia, \newline melitis & cardiovascular, \newline nonalcoholic, \newline obesity, \newline mellitus, \newline polycystic & hypertension, \newline obesity, \newline arthritis, \newline cancer, \newline alzheimer & diabetics, \newline  hypertension, \newline diabetic, \newline diabetes\_mellitus, \newline  heart\_disease \\
\cline{2-6}
 & peptic ulcer disease & scleroderma, \newline duodenal, \newline crohn, \newline gastroduodenal, \newline diverticular & gastritis, \newline alcoholism, \newline rheumatic, \newline ischaemic, \newline nephropathy & ulcers, \newline arthritis, \newline diseases, \newline diabetes, \newline stomach & ichen\_planus, \newline Candida\_infection, \newline vaginal\_yeast\_infections, \newline oral\_thrush, \newline dermopathy  \\
 \cline{2-6}
 & colon cancer & breast, \newline ovarian, \newline prostate, \newline postmenopausally, \newline caner & breast, \newline mcf, \newline cancers, \newline tumor\_suppressing, \newline downregulation & breast, \newline prostate, \newline cancers, \newline tumor, \newline liver & breast, \newline prostate, \newline tumor, \newline pre\_cancerous\_lesion, \newline  cancerous\_polyp  \\
\hline
\multirow{3}{*}{Symptom}  & dyspnea & palpitations, \newline orthopnea, \newline exertional, \newline doe, \newline dyspnoea & sweats, \newline orthopnea, \newline breathlessness, \newline hypotension, \newline rhonchi & shortness, \newline breathlessness, \newline cyanosis, \newline photophobia, \newline faintness & dyspnoea, \newline pruritus, \newline nasopharyngitis, \newline symptom\_severity, \newline rhinorrhea \\
 \cline{2-6}
  & sore throat & scratchy, \newline thoat, \newline cough, \newline runny, \newline thraot & runny, \newline rhinorrhea, \newline myalgia, \newline swab\_fecal, \newline nose & shoulder, \newline stomach, \newline nose, \newline chest, \newline neck & soreness, \newline bruised, \newline inflammed, \newline contusion, \newline sore\_triceps \\
 \cline{2-6}
  & low blood pressure & readings, \newline pressue, \newline presssure, \newline bptru, \newline systolically & dose, \newline cardio\_ankle, \newline ncbav, \newline preload, \newline gr & because, \newline result, \newline high, \newline enough, \newline higher & splattering\_tombstones, \newline Zapping\_nerves\_helps, \newline pressue, \newline Marblehead\_Swampscott\_VNA, \newline pill\_Norvasc \\
  \hline
\multirow{2}{*}{Drug}  & opioid & opiate, \newline benzodiazepine, \newline opioids, \newline sedative, \newline polypharmacy & opioids, \newline nmda\_receptor, \newline affective\_motivational, \newline naloxone\_precipitated, \newline hyperlocomotion & analgesic, \newline opiate, \newline opioids, \newline anti-inflammatory, \newline analgesics & opioids, \newline opioid\_analgesics, \newline opioid\_painkillers, \newline antipsychotics, \newline tricyclic\_antidepressants  \\
 \cline{2-6}
  & aspirin & ecotrin, \newline uncoated, \newline nonenteric, \newline effient, \newline onk & chads, \newline vasc, \newline newer, \newline cha, \newline angina & ibuprofen, \newline tamoxifen, \newline pills, \newline statins, \newline medication & dose\_aspirin, \newline ibuprofen, \newline statins, \newline statin, \newline  calcium\_supplements \\
\hline
\end{tabular}
\end{table}

\begin{figure}[!h]
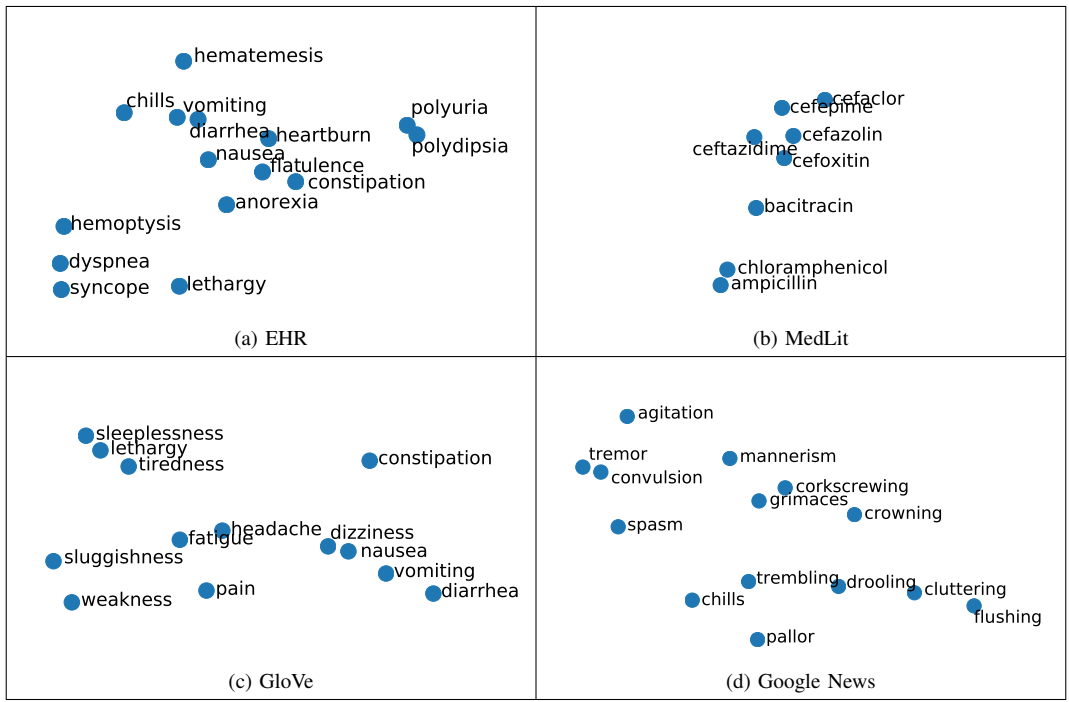

\centering
\begin{tabular}{|c|c|}
\hline
\subfloat[EHR]{\label{fig.clusters.a}\includegraphics[width=0.4\textwidth]{echcn1_sub.pdf}} &
\subfloat[MedLit]{\label{fig.clusters.b}\includegraphics[width=0.23\textwidth]{pmc_sub.pdf}} \\
\hline
\subfloat[GloVe]{\label{fig.clusters.c}\includegraphics[width=0.4\textwidth]{wiki_sub.pdf}} &
\subfloat[Google News]{\label{fig.clusters.d}\includegraphics[width=0.4\textwidth]{google_sub.pdf}} \\
\hline
\end{tabular}
\caption{Examples of word clusters in the visualization of word embeddings trained from four corpora using t-SNE.}
\label{fig.clusters}
\end{figure}

In order to show different aspects of medical concepts captured by word embeddings trained from different corpora, we extracted 377 medical terms from the UMNSRS dataset \cite{pakhomov2010semantic,pakhomov2016corpus} and visualized the word embeddings for these medical terms in a two-dimensional plot using t-distributed stochastic neighbor embedding (t-SNE) \cite{maaten2008visualizing}. Example clusters of medical terms in the word embeddings are shown in Figure \ref{fig.clusters}. Figure \ref{fig.clusters.a} depicts a cluster of symptoms, such as \textit{heartburn}, \textit{vomiting} and \textit{nausea}, from the word embeddings trained on EHR. Figure \ref{fig.clusters.b} shows a cluster of antibiotic medications, such as \textit{bacitracin}, \textit{cefoxitin}, and \textit{chloramphenicol}, based on MedLit embeddings. Figures \ref{fig.clusters.c} and \ref{fig.clusters.d} illustrate clusters of symptoms from the GloVe and Google News embeddings, respectively. Since we did not employ any clustering method, these clusters were intuitively observed from the two-dimensional plot. The visualization of the entire set of 377 medical terms using word embeddings trained from four different corpora is provided in the supplementary file. 

\section{Quantitative Evaluation}

We conducted both extrinsic and intrinsic quantitative evaluation, where the former used four published datasets for measuring semantic similarity between medical terms and the latter used downstream biomedical NLP tasks to evaluate word embeddings.

\subsection{Intrinsic Evaluation}

We tested word embeddings on four published biomedical measurement datasets commonly used to measure semantic similarity between medical terms. The first is Pedersen's dataset \cite{pedersen2007measures} that consists of 30 medical term pairs that were scored by physician experts according to their relatedness. The second is Hliaoutakis's dataset \cite{hliaoutakis2005semantic} consisting of 34 medical term pairs with similarity scores obtained by human judgments. The third, the MayoSRS dataset developed by Pakhomov et al. \cite{pakhomov2011towards}, consists of 101 clinical term pairs whose relatedness was determined by nine medical coders and three physicians from Mayo Clinic. The relatedness of each term pair was assessed based on a four point scale: (4.0) practically synonymous, (3.0) related, (2.0) marginally related and (1.0) unrelated. We evaluated the word embeddings using the mean score of the physicians and medical coders. The fourth, UMNSRS similarity dataset developed by Pakhomov et al. \cite{pakhomov2010semantic}, consists of 566 medical term pairs whose semantic similarity was determined independently by eight medical residents from the University of Minnesota Medical School. The similarity and relatedness of each term pair was annotated based on a continuous scale by having the resident touch a bar on a touch sensitive computer screen to indicate the degree of similarity or relatedness. 

For each pair of medical terms in the testing datasets, we used Equations (\ref{equ.simi1}) and (\ref{equ.simi2}) to calculate the semantic similarity for each pair. Since some medical terms may not exist in the vocabulary of word embeddings, we used fastText \cite{bojanowski2016enriching} to compute word vectors for these out-of-vocabulary (OOV) medical terms. Specifically, we built character n-gram vectors analogous to fastText's output by converting each word (e.g., ``abcdef'') in the word embeddings to 3-gram (i.e., trigram) format (i.e., ``abc'', ``bcd'', ``cde'', ``def'') with vector representation of each trigram the same as that of the original word. After converting all the words, we utilized the averaged vector for the identical trigram extracted from different words (e.g., the vector for ``abcdef'' is $\theta_1$ and that for ``defg'' is $\theta_2$, the final vector for trigram ``def'' is $\frac{1}{2}(\theta_1+\theta_2)$ since ``def'' is a shared trigram between the two words). Since each word with the number of characters greater than or equal to 3 can be represented as a bag of character trigrams, fastText represents an OOV medical term as the normalized sum of the vector representations of its trigrams \cite{bojanowski2016enriching}. The Pearson correlation coefficient was employed to calculate the correlation between similarity scores from human judgments and those from word embeddings. 

Table \ref{tab.similarity} lists the Pearson correlation coefficient results for the four datasets. Overall, the semantic similarity captured by the word embeddings trained on EHR are closer to human experts' judgments, compared with other word embeddings. MedLit performs worse than EHR but has a comparative result for the UMNSRS dataset. GloVe and Google News are inferior to EHR and MedLit, and perform similarly in representing medical semantics. Note that the four datasets and corresponding semantic similarity scores from both human experts and word embeddings are provided in the supplementary Excel file.

\begin{table}[!h]
\centering
\footnotesize
\caption{Pearson correlation coefficient between similarity scores from human judgments and those from word embeddings on four measurement datasets. The asterisk indicates that difference between word embeddings trained on EHR and those on other resources is statistically significant using t-test (p$<$0.01).}
\label{tab.similarity}
\begin{tabular}{ccccc}
\hline
Dataset & EHR & MedLit & GloVe & Google News \\
\hline
Pedersen's & 0.632* & 0.569 & 0.403 & 0.357 \\
Hliaoutakis's & 0.482* & 0.311 & 0.247 & 0.243 \\
MayoSRS & 0.412* & 0.300 & 0.082 & 0.084 \\
UMNSRS & 0.440* & 0.404 & 0.177 & 0.154 \\
\hline
\end{tabular}
\end{table}

\subsection{Extrinsic Evaluation}

Extrinsic evaluations are used to measure the impact of word embeddings to specific biomedical NLP tasks. In this evaluation, we tested the word embeddings on three biomedical NLP tasks, namely clinical IE, biomedical IR, and RE. 

\subsubsection{Clinical Information Extraction}
Two clinical IE tasks were utilized to evaluate the word embeddings. The first task is an institutional task while the second is a shared task. Using the first task, we would like to examine whether the word embeddings trained on our institutional corpus perform better than external pre-trained word embeddings on a local institutional IE task. We also would like to investigate whether the results are consistent on a global shared task.


In the first experiment, we evaluated the word embeddings on an institutional IE task at Mayo Clinic. In this task, a set of 1000 radiology reports was given to detect whether a hand and figure/wrist fracture could be identified. Reports were drawn from a cohort of residents of Olmsted County, aged 18 or older, who experienced fractures in 2009-2011. Each report was annotated by a medical expert with multiple years of experience abstracting fractures by assigning ``1'' if a hand and figure/wrist fracture was found, or ``0'' otherwise. 

In our experiment, the word embeddings were employed as features for machine learning models and evaluated by precision, recall, and F1 scores \cite{wang2018distant}. For a clinical document $d=\{w_1,w_2,..,w_M\}$ where $w_i, i=1,2,...,M$ is the $i$th word and $M$ is the total number of words in this document, the feature vector $\mathbf{x}$ of document $d$ is defined by
$$\mathbf{x}=\frac{1}{M} \sum_{i}^{M}\mathbf{x}_i,$$
where $\mathbf{x}_i$ is the embedding vector for word $w_i$ from the word embedding matrix. Then $\mathbf{x}$ was utilized as input to a conventional machine learning model, which is Support Vector Machine (SVM) in this experiment. We performed 10-fold cross validation on the dataset. The means of precision, recall, and F1 scores from the 10-fold cross validation was reported, which are defined below: 
$$Precision = \frac{1}{10}\sum_i Precision_i=\frac{1}{10}\sum_i \frac{TP_i}{TP_i+FP_i},$$
$$Recall=\frac{1}{10}\sum_i Recall_i=\frac{1}{10}\sum_i \frac{TP_i}{TP_i+FN_i},$$
$$F1\text{ }score=\frac{1}{10}\sum_i F1\text{ }score_i=\frac{1}{10}\cdot\sum_i \frac{2TP_i}{2TP_i+FP_i+FN_i},$$
where TP, TN, FP, and FN represent true positives, true negatives, false positives, and false negatives, respectively, and $i={1,2,...,10}$ represents the $i$th fold cross validation. As a comparison, the baseline method used term frequency features as input. 

The experimental results are listed in Table \ref{tab.localie}. The word embeddings trained on EHR are superior to other word embeddings in terms of all metrics (precision: 0.974, recall: 0.972, F1 score: 0.972) with statistical significance using t-test (p$<$0.01). The fracture dataset in this experiment is curated from the same EHR system as the EHR corpus used to train word embeddings, and thus they have identical sublanguage characteristics. The word embeddings trained on MedLit also have comparable results (precision: 0.946, recall: 0.943, F1 score: 0.942). Since this task is a medical task with specific medical terminologies, the word embeddings trained on Google News have the worst performance. However, the word embeddings trained on GloVe are close to those trained on EHR with 0.02 difference on F1 score without statistical significance (p$<$0.01). This experiment shows that word embeddings trained on a local corpus have the best performance for a local task but those trained on an external Wikipedia corpus also have comparable performance. 

\begin{table}[!h]
\centering
\footnotesize
\caption{Results of the institutional fracture extraction task using word embeddings trained from four different corpora. The asterisk indicates that the difference between word embeddings trained on EHR and those on other resources is statistically significant using t-test (p$<$0.01).}
\label{tab.localie}
\begin{tabular}{cccccc}
\hline
Metric & baseline & EHR & MedLit & GloVe & Google News \\
\hline
Precision & 0.612 & 0.974* & 0.946 & 0.951 & 0.809 \\
Recall & 0.612 & 0.972* & 0.943 & 0.950 & 0.856 \\
F1 score & 0.609 & 0.972* & 0.942 & 0.950 & 0.823 \\
\hline
\end{tabular}
\end{table}

Secondly, we tested the word embeddings on the 2006 i2b2 (Informatics for Integrating Biology to the Bedside) smoking status extraction shared task \cite{uzuner2008identifying}. Participants of this task were asked to develop automatic NLP systems to determine the smoking status of patients from their discharge records in Partners HealthCare. For each discharge record, an automatic system should be able to categorize it into five pre-determined smoking status categories: past smoker, current smoker, smoker, non-smoker, and unknown, where a past and a current smoker are distinguished based on temporal expressions in the patient's medical records. The dataset contains a total of 389 documents, including 35 documents of current smoker, 66 of non-smoker, 36 of past smoker, and 252 of unknown. The settings of this shared task are identical to those of the previous local institutional IE task: SVM was utilized as the machine learning model; 10-fold cross validation was performed; term frequency features were used as input in the baseline; and the means of precision, recall and F1 scores were obtained as metrics.

The experimental results are shown in Table \ref{tab.ie}. First, it is obvious that the word embedding features perform better than term frequency features due to the semantics embedded in word embeddings, which is consistent with the previous local institutional IE task. The word embeddings trained on EHR produced the best performance with a F1 score of 0.900. The reason might be that the smoking dataset has the similar sublanguage characteristics as the EHR corpus. This result indicates that the effective word embeddings can be shared across institutions for clinical IE tasks. Another interesting observation is that the performance of word embeddings trained on Google News is close to that trained on EHR corpus with a comparable F1 score and a better recall. The performance difference is not statistically significant (p$<$0.01). This implies that word embeddings trained on a public dataset may not be definitely inferior to these trained on a medically specific dataset for a medical IE task. The likely cause is that the terminology used in the smoking status extraction task also appears frequently in the news, such as medications and advice for smokers.

\begin{table}[!h]
\centering
\footnotesize
\caption{Results of the i2b2 2006 smoking status extraction task using word embeddings trained from four different corpora.}
\label{tab.ie}
\begin{tabular}{cccccc}
\hline
Metric & baseline & EHR & MedLit & GloVe & Google News \\
\hline
Precision & 0.692 & \textbf{0.919} & 0.878 & 0.893 & 0.910 \\
Recall & 0.486 & 0.903 & 0.871 & 0.889 & \textbf{0.905} \\
F1 score & 0.539 & \textbf{0.900} & 0.867 & 0.884 & 0.897 \\
\hline
\end{tabular}
\end{table}

\subsubsection{Biomedical Information Retrieval}

To evaluate word embeddings for biomedical IR, we utilized the dataset provided by the Text REtreival Conference 2016 Clinical Decision Support (TREC 2016 CDS) track. The TREC 2016 CDS track focuses on biomedical literature retrieval that helps physicians find the precise literature information and make the best clinical decision at the point of care \cite{roberts2016overview}. The query topics were generated from EHRs in the MIMIC-III dataset \cite{johnson2016mimic}. Those topics were categorized into three most common types, \textit{Diagnosis}, \textit{Test} and \textit{Treatment}, according to physicians' information needs, and 10 topics were provided for each type. Each topic is comprised of a \textit{note} field (admission note), a \textit{description} field (jargons and clinical abbreviations are removed) and a \textit{summary} field (simplified version of the description). The participants were required to use only one of these three fields in their submissions and at least one submission must utilize the \textit{note} field. Submitted systems should retrieve relevant biomedical articles from a given PMC article collection for each given query topic to answer three corresponding clinical questions: \textit{What is the patient's diagnosis? What tests should the patient receive? How should the patient be treated?}. Each IR system can retrieve up to 1000 documents per query.

In order to make the comparison as fair as possible, we first implemented a simple IR system as the baseline system using the original queries following the study in \cite{wang2016ensemble}, and then employed the simplest query expansion method using the word embeddings. We used the \textit{summary} field in the query, removed the stop words, and expanded each of the left query terms with five most similar terms from word embeddings. Take the query ``A 78 year old male presents with stools and melena'' as an example, the term ``male'' was expanded by ``female gentleman lady caucasian man'', ``stools'' by ``stooling liquidy voluminous semiformed tenesmus'', and ``melena'' by ``hematemesis hematochezia melana brbpr hematemasis''. We assigned weight 0.8 to the original query and 0.2 to the expanded query.  Indri \cite{strohman2005indri} was utilized as our indexing and retrieval tool. The preprocessing for the corpus included stopword removal and Porter stemming. The stopword list was based on the MedLit stopwords \footnote{http://www.ncbi.nlm.nih.gov/books/NBK3827/table/pubmedhelp.T.stopwords/}. The \textit{article-id}, \textit{title}, \textit{abstract}, and \textit{body} fields of each document in the corpus were indexed. Language models with two-stage smoothing \cite{zhai2002two} was used to obtain all the retrieval results. Four official metrics, namely Inferred Normalized Discounted Cumulated Gain (infNDCG)\cite{yilmaz2008simple}, Inferred Average Precision (infAP)\cite{yilmaz2008simple}, Precision at 10 (P@10), and Mean Average Precision (MAP), were utilized to measure the IR performance. infNDCG measures the document ranking quality of an IR system; infAP measures the retrieval effectiveness given incomplete judgments for an IR system; P@10 is the number of relevance documents among the top 10; and MAP is the mean of the average precision scores for each query in a set of queries.

Table \ref{tab.ir} lists the results of using the word embeddings trained from different resources for query expansion on the TREC 2016 CDS track. It is interesting that the word embeddings based query expansion method failed to improve the retrieval performance, and even worsened the performance when infAP and MAP were metrics. By comparing the retrieval performance, we observe that EHR and MedLit perform slight better than GloVe and Google News without statistical significance (p$<$0.01). This result implies that applying word embeddings trained from different resources has no significant improvement for the biomedical IR task. 

\begin{table}[!h]
\centering
\footnotesize
\caption{Information retrieval results of using word embeddings trained from four different corpora for query expansion on the TREC 2016 CDS track.}
\label{tab.ir}
\begin{tabular}{cccccc}
\hline
Metric & baseline & EHR & MedLit & GloVe & Google News \\
\hline
infNDCG & 0.249 & \textbf{0.250} & 0.249 & 0.249 & 0.238 \\
infAP & \textbf{0.058} & 0.056 & 0.055 & 0.051 & 0.052 \\
P@10 & 0.247 & 0.243 & \textbf{0.248} & 0.233 & 0.243 \\
MAP & \textbf{0.067} & 0.063 & 0.065 & 0.063 & 0.059 \\
\hline
\end{tabular}
\end{table}

\subsubsection{Relation Extraction}

For the RE task, we considered drug-drug interaction (DDI) extraction, which is a specific RE task in the biomedical domain. DDI is an unexpected change in a drug's effect on the human body when the drug and a second drug are co-prescribed and taken together. Automatically extracting DDI information from literature is a challenging and important research topic since the volume of the published literature grows rapidly and greatly. In this experiment, we evaluated the word embeddings on the DDIExtraction 2013 challenge corpus \cite{segura2013semeval}. The dataset for DDIExtraction 2013 was composed of sentences describing DDIs from the DrugBank database and MedLine abstracts. In this dataset, drug entities and DDIs were annotated at the sentence level and each sentence could contain two or more drugs. An RE system should be able to automatically extract DDI drug pairs from a sentence. We exploited the baseline system introduced in \cite{wang2017dependency} where features include words and word bigrams with binary values indicating their presence or absence in a sentence, cosine similarity between centroid vector of each class and the instance, negation (three features indicating negation before the first main drug, between two main drugs, and after the two main drugs). We concatenated the word embeddings to the baseline features and tested the performance. Since Random Forest \cite{breiman2001random} has the best performance in \cite{wang2017dependency}, we utilized it as the classifier with 10-fold cross validation.

Table \ref{tab.re} shows the F1 scores of Random Forest using word embeddings trained from different resourceson the DDIExtraction 2013 challenge. We can see that the overall performance of word embeddings trained on Google News is the best. The reason is that the semantics of general English terms in the context of drug mentions are more important for determining the drug interactions. For example, in the sentence ``Acarbose may interact with metformin'', the term ``interact'' is crucial to classify the relation. Since these crucial terms are generally not medical terminology, word embeddings trained on Google News where the corpus represents general English are able to capture the semantics of these terms. However, Google News outperformed other resources but not conclusive in statistical significance using t-test (p$<$0.01). Another interesting observation is that word embeddings trained from MedLit have the best performance for the DrugBank corpus while these from Google News perform the best for the MedLine corpus. Though MedLine abstracts are from MedLit articles, this result shows that word embeddings trained from the same corpus are not necessarily superior to other embeddings. 


\begin{table}[!h]
\centering
\footnotesize
\caption{F1 scores of the DDIExtraction 2013 challenge using word embeddings trained from four different corpora.}
\label{tab.re}
\begin{tabular}{cccccc}
\hline
Category & baseline & EHR & MedLit & GloVe & Google News \\
\hline
DrugBank (5265 pairs) & 0.590 &  0.708 &  \textbf{0.715} & 0.714 & 0.705  \\
MedLine (451 pairs) & 0.690  &  0.696 &  0.690 & 0.699 & \textbf{0.708}  \\
Total (5716 pairs) & 0.760 &  0.789 & 0.788  & 0.787 & \textbf{0.790}  \\
\hline
\end{tabular}
\end{table}

\section{Conclusion and Discussion}\label{sec.con}

In this study, we provide an empirical evaluation of word embeddings trained from four different corpora, namely clinical notes, biomedical publications, Wikipedia, and news. We performed the evaluation qualitatively and quantitatively. For the qualitative evaluation, we selected a set of medical words and impressionistically evaluated the five most similar medical words. We then analyzed word embeddings through the visualization of those word embeddings. We conducted both extrinsic and intrinsic evaluation for the quantitative evaluation. The intrinsic evaluation directly tested  semantic relationships between medical words using four published datasets for measuring semantic similarity between medical terms while the extrinsic evaluation evaluated word embeddings in three downstream biomedical NLP applications, i.e., clinical IE, biomedical IR, and RE.

Based on the evaluation results, we can draw the following conclusions. First, the word embeddings trained on EHR and MedLit can capture the semantics of medical terms better than those trained on GloVe and Google News, and find more relevant similar medical terms. However, EHR finds similar terms vis a vis clinical language while MedLit contains more medical terminology used in medical articles, and finds similar words mostly from a medical research perspective. Second, the medical semantic similarity captured by the word embeddings trained on EHR and MedLit are closer to human experts' judgments, compared to these trained on GloVe and Google News. Third, there does not exist a consistent global ranking of word embeddings for the downstream biomedical NLP applications. However, adding word embeddings as extra features will improve results on most downstream tasks. Finally, word embeddings trained from biomedical domain corpora do not necessarily have better performance than those trained on other general domain corpora. That is, there might be no significant difference when word embeddings trained from an out-domain corpus are employed for a biomedical NLP application. However, the performance of word embeddings trained from a local institutional corpus might perform better for local institutional NLP tasks.

Our experiments implicitly show that applying word embeddings trained from corpora in a general domain, such as Wikipedia and news, is not significantly inferior to applying those obtained from biomedical or clinical domain, which is usually difficult to access due to privacy. This result is consistent with but more general than the conclusion drawn in \cite{pakhomov2016corpus}. Thus, a lack of access to a domain-specific corpus is not necessarily a barrier for the use of word embeddings in practical implementations. 

As a future direction, we would like to evaluate word embeddings on more downstream biomedical NLP applications, such as medial named entity recognition and clinical note summarization. We will investigate whether word embeddings trained from different resources represent language characteristics differently for a corpus, such as term frequency and medical concepts. We also want to assess word embeddings across health care institutions using different EHR systems and investigate how sublanguage characteristics affect the portability of word embeddings. Moreover, we want to apply clustering methods on word embeddings and compare the word-level and concept-level difference between clusters of medical terms. 

There are a few limitations in this study. First, we only examined the word embeddings trained on the EHR from Mayo Clinic, which might have introduced bias into the conclusion as the EHR quality may vary by institutions. However, it is challenging to obtain word embeddings trained on EHR data from multiple sites. We are currently exploring the use of privacy-preserving techniques for obtaining embeddings from multiple sites leveraging our prior work \cite{huang2018privacy} to have more generalizable embeddings. Second, we tested only two widely used public pre-trained word embeddings. There are a number of word embeddings publicly available\footnote{https://github.com/3Top/word2vec-api}. Third, the generalizability of the results for the biomedical IR and RE tasks may be questionable since we only used one shared task dataset for each task to evaluate the word embeddings.

\section{Acknowledgement}\label{sec.ack}

This work has been supported by the National Institute of Health (NIH) grants R01LM011934, R01GM102282, and U01TR002062.

\bibliographystyle{IEEEtran}
\bibliography{dl4ehr}

\appendices

\section{Intrinsic evaluation of word embeddings with different dimensions.}\label{ap.1}

Table \ref{tab.append.similarity} shows the intrinsic evaluation results of word embeddings using different dimensions. 

\begin{table}[!h]
\centering
\footnotesize
\caption{Pearson correlation coefficient between the similarity scores computed by word embeddings using different dimensions (d) and those assigned by human experts on four datasets.}
\label{tab.append.similarity}
\begin{tabular}{cp{1cm}p{1cm}p{1cm}p{1cm}p{1cm}p{1cm}p{1cm}p{1cm}p{1.7cm}}
\hline
Dataset & EHR (d=20) & EHR (d=60) & EHR (d=100) & MedLit (d=20) & MedLit (d=60) & MedLit (d=100) & GloVe (d=50) & GloVe (d=100) & Google News (d=300) \\
\hline
Pedersen's & 0.390 & 0.542 & 0.632 & 0.304 & 0.569 & 0.363 & 0.334  &  0.403 & 0.357 \\
Hliaoutakis's & 0.333 & 0.417 &  0.482 & 0.117 &  0.311 & 0.164 & 0.159 &   0.247 & 0.243 \\
MayoSRS & 0.192 & 0.296 &  0.412 & 0.177 &  0.300 & 0.154 & 0.001  &  0.082 & 0.084 \\
UMNSRS & 0.310 & 0.375 &  0.440 & 0.295 &  0.404 & 0.396 & 0.190  &  0.177 & 0.154 \\
\hline
\end{tabular}
\end{table}

%
%
%
%
%

\end{document}


%
\title{Supplementary File}

\maketitle


\section{Visualization of word embeddings.}\label{ap.2}

\begin{figure}
\centering
\includegraphics[width=0.9\textwidth]{echcn1.pdf}
\caption{Visualization of word embeddings trained on EHR.}
\label{fig.ehr}
\end{figure}

\begin{figure}
\centering
\includegraphics[width=0.9\textwidth]{pmc.pdf}
\caption{Visualization of word embeddings trained on MedLit.}
\label{fig.pmc}
\end{figure}

\begin{figure}
\centering
\includegraphics[width=0.9\textwidth]{wiki.pdf}
\caption{Visualization of word embeddings trained on GloVe.}
\label{fig.wiki}
\end{figure}

\begin{figure}
\centering
\includegraphics[width=0.9\textwidth]{google.pdf}
\caption{Visualization of word embeddings trained on Google News.}
\label{fig.google}
\end{figure}
